
\magnification=1200
\baselineskip=13pt
\overfullrule=0pt
\tolerance=100000
 at 8truept

\nopagenumbers
{\hfill \hbox{\vbox{\settabs 1\columns
\+ UR-1424 \cr
\+ ER-40685-873\cr
\+ hep-th/9506096\cr
}}}
\bigskip
\bigskip
\baselineskip=15pt

\centerline{\bf The Supersymmetric Two Boson Hierarchy\footnote{\rm *}{\rm
Invited talk given by A.D. at the Workshop in Theoretical and Mathematical
Physics, CAM'95, University of Laval, Quebec, Canada, June 11--16, 1995.}}
\vfill

\centerline{J. C. Brunelli}
\medskip
\centerline{and}
\medskip
\centerline{Ashok Das}
\medskip
\medskip
\centerline{Department of Physics and Astronomy}
\centerline{University of Rochester}
\centerline{Rochester, NY 14627, USA}
\vfill

\centerline{\bf {Abstract}}

\medskip
\medskip

We summarize all the known properties of the supersymmetric
integrable Two Boson equation. We present its nonstandard Lax formulation and
tri-Hamiltonian structure, its reduction to the supersymmetric nonlinear
Schr\"odinger equation and the local as well as nonlocal conserved charges.
We also present the algebra of the conserved charges and identify
its second Hamiltonian structure with the twisted $N=2$
superconformal algebra.
\vfill
\eject
\headline={\hfill\folio}
\pageno=1
\bigskip
\noindent {\bf 1. {Introduction}}
\medskip

The study of integrable models has provided for a long time a crossover arena
between mathematical and theoretical physics [1]. Recently, these models have
found a relevant role in the study of strings through the matrix models [2].
So, its is natural to study supersymmetric integrable models since they are
likely to play an important role in the superstrings [3].

The most widely studied supersymmetric integrable system is the supersymmetric
KdV (sKdV) equation [4]. But, there are other integrable systems, which can be
obtained by supersymmetrization of other well known bosonic ones and may have
an important role in physical applications in the study of the
super matrix models. In this talk we give a description of some of our recent
results [5-10] on the
supersymmetric Two Boson (sTB) equation [6]. A detailed review of this
system and other results not covered in this talk can be found in our review
paper [10].
\bigskip
\noindent {\bf 2. {Two Boson Equation}}
\medskip

The Two Boson system is a dispersive generalization of the long water wave
equation [11] and has appeared in the literature in the study of bosonic matrix
models [12]. The equations for this $1+1$ dimensional integrable system are
$$
\eqalign{
{\partial  J_0 \over \partial  t} &= (2 J_1 + J_0^2 -
J^\prime_0 )^\prime\cr
\noalign{\vskip 4pt}%
{\partial  J_1 \over \partial  t} &= (2 J_0 J_1 +
J^\prime_1 )^\prime\cr
}\eqno(1)
$$
This equation can be obtained from the Lax operator
$$
L =  \partial  - J_0 + \partial^{-1} J_1 \eqno(2)
$$
through a nonstandard Lax equation [11,13]
$$
{\partial  L \over \partial  t} = \left[ L, \left( L^2 \right)_{\geq 1}
\right] \eqno(3)
$$
where $()_{\geq 1}$ refers to the differential part of the pseudo-differential
operator. The conserved charges of the system are in involution and can be
obtained in the conventional manner from
$$
H_n = \ {\rm Tr}\ L^n = \int dx\ {\rm Res}\ L^n \qquad \qquad
n = 1,2,3,\dots \eqno(4)
$$
\eject
The TB equation has a tri-Hamiltonian structure [11] which can be obtained from
modified Gelfand-Dikii brackets [13]. The second Hamiltonian structure is
related to a Virasoro-Kac-Moody algebra for a $U(1)$ current (an affine
algebra) [13]. Also, under an appropriate field redefinitions or reductions we
can obtain from it the well known integrable systems such as the KdV, mKdV
and the nonlinear Schr\"odinger (NLS) equations.

\bigskip
\noindent {\bf 3. {Supersymmetric Two Boson Equation}}
\medskip

A supersymmetric generalization of (1) can be constructed [6] if we introduce
the fermionic superfields
$$
\eqalign{\Phi_0 &= \psi_0 + \theta J_0\cr
\Phi_1 &= \psi_1 + \theta J_1 \cr
}
\eqno(5)
$$
and the supercovariant derivative
$$
D = {\partial  \over \partial  \theta} + \theta {\partial
 \over \partial  x}
\eqno(6)
$$
We use the notation [6] where $z=(x,\theta)$ defines the coordinates of
the superspace with $\theta$ representing the Grassmann coordinate.

The most general, local, dynamical equations in the superspace which are
consistent with all the canonical dimensions and which reduce to (1) in the
bosonic limit are
$$
\eqalign{
{\partial  \Phi_0 \over \partial  t} = &-(D^4 \Phi_0) + 2(D \Phi_0)
(D^2 \Phi_0) + 2(D^2 \Phi_1)\cr
&+ a_1 D(\Phi_0 (D^2 \Phi_0)) + a_2 D (\Phi_0 \Phi_1)\cr
\noalign{\vskip 4pt}%
{\partial  \Phi_1 \over \partial  t} = &(D^4 \Phi_1) + b_1 D( ( D^2
\Phi_1 ) \Phi_0) + 2(D^2 \Phi_1)(D \Phi_0 ) - b_2 D(\Phi_1
(D^2 \Phi_0))\cr
& +2(D \Phi_1 )(D^2 \Phi_0) + b_3 \Phi_1 \Phi_0(D^2 \Phi_0 ) + b_4 D(\Phi_1
\Phi_0) (D \Phi_0)\cr
&+ b_5 D(\Phi_0 (D^4 \Phi_0)) +b_6 D(\Phi_0 (D^2 \Phi_0)) (D \Phi_0)\cr
}
\eqno(7)
$$
where $a_i$ and $b_i$ are arbitrary parameters. However, equations (7) are
integrable only for  specific choices of $a_i$ and $b_i$. In fact, a
consistent Lax representation can be obtained for the system (7) with the Lax
operator
$$
L = D^2 - (D \Phi_0) + D^{-1} \Phi_1
\eqno(8)
$$
and the nonstandard Lax equation
$$
{\partial  L \over \partial  t} = \left[ L, (L^2)_{\geq 1}
\right]\eqno(9)
$$
where $D^{-1}=\partial^{-1}D$. Here, (8)
reduces in the bosonic limit to (2).  So,
the most general supersymmetric extension of the dynamical equations (1) which
is integrable is given by
$$
\eqalign{{\partial  \Phi_0 \over \partial  t} &=
 - (D^4 \Phi_0) + (D(D\Phi_0)^2)+ 2(D^2 \Phi_1)\cr
\noalign{\vskip 4pt}%
{\partial  \Phi_1 \over \partial  t} &=
 (D^4 \Phi_1) + 2(D^2((D \Phi_0) \Phi_1))\cr
}
\eqno(10)
$$
These are the supersymmetric Two Boson equations (sTB) [6]  with the
nonstandard Lax
representation given by (8) and (9). In components the equations (10) read
$$
\eqalign{
{\partial  J_0 \over \partial  t} &= ( 2J_1 + J_0^2 - J_0^\prime)^\prime\cr
\noalign{\vskip 4pt}%
{\partial  \psi_0 \over \partial  t} &= 2 \psi_1^\prime + 2 \psi_0^\prime
J_0 - \psi_0^{\prime \prime}\cr
\noalign{\vskip 4pt}%
{\partial  J_1 \over \partial  t} &= ( 2J_0 J_1 + J_1^\prime
 + 2 \psi_0^\prime \psi_1)^\prime\cr
\noalign{\vskip 4pt}%
{\partial  \psi_1 \over \partial  t} &= ( 2 \psi_1
J_0 + \psi_1^\prime)^\prime\cr
}
\eqno(11)
$$
and it is straightforward to check that these equations are invariant under the
$N=1$ supersymmetry transformation
$$
\eqalign{
\delta  J_0 &= \epsilon \psi_0^\prime\cr
\delta  J_1 &= \epsilon \psi_1^\prime\cr
\delta \psi_0 &= \epsilon J_0\cr
\delta  \psi_1 &= \epsilon J_1\cr
}
\eqno(12)
$$
In fact, as we will see, there is another supersymmetry transformation as well
as additional symmetries present in our system (10).

Since equations (10) are integrable we have an infinite number of local
conserved charges in involution which can be obtained from
$$
Q_n=\hbox{sTr}\,L^n=\int dz\,\hbox{sRes}\,L^n\qquad n=1,2,3,\dots \eqno(13)
$$
The first few charges are
$$
\eqalign{
Q_1=&-\int dz\, \Phi_1\cr
Q_2=&2\int dz\, (D\Phi_0)\Phi_1\cr
Q_3=&3\int dz\,\Bigl[(D^3\Phi_0)-(D\Phi_1)-(D\Phi_0)^2\Bigr]\Phi_1\cr
Q_4=&2\int
dz\,\Bigl[2(D^5\Phi_0)+2(D\Phi_0)^3+6(D\Phi_0)(D\Phi_1)-
3\left(D^2(D\Phi_0)^2\right)
\Bigr]\Phi_1\cr
}\eqno(14)
$$
They are bosonic and are invariant under the supersymmetry transformation (12).

Defining the Hamiltonians of the sTB system as
$$
H_n={(-1)^{n+1}\over n}Q_n\eqno(15)
$$
we can write the sTB equations (10) as a Hamiltonian system [6]
$$
\partial_t\pmatrix{\Phi_0\cr
\noalign{\vskip 10pt}%
\Phi_1}
={\cal D}_1
\pmatrix{{\delta H_{3}\over\delta\Phi_0}\cr
\noalign{\vskip 10pt}%
{\delta H_{3}\over\delta\Phi_1}}=
{\cal D}_2
\pmatrix{{\delta H_{2}\over\delta\Phi_0}\cr
\noalign{\vskip 10pt}%
{\delta H_{2}\over\delta\Phi_1}}=
{\cal D}_3
\pmatrix{{\delta H_{1}\over\delta\Phi_0}\cr
\noalign{\vskip 10pt}%
{\delta H_{1}\over\delta\Phi_1}}
\eqno(16)
$$
where the first structure has the local form
$$
{\cal D}_1=\pmatrix{0 & -D\cr
\noalign{\vskip 5pt}%
-D & 0}\eqno(17)
$$
and the second has the  nonlocal form [18]
$$
{\cal D}_2=\pmatrix{-2D-2D^{-1}\Phi_1D^{-1}+D^{-1}(D^2\Phi_0)D^{-1}&
D^3-D(D\Phi_0)+D^{-1}\Phi_1D\cr
\noalign{\vskip 20pt}%
-D^3-(D\Phi_0)D-D\Phi_1D^{-1}&-\Phi_1D^2-D^2\Phi_1}\eqno(18)
$$
Defining
$$
R={\cal D}_2{\cal D}_1^{-1}\eqno(19)
$$
the third structure, which is highly nonlocal, can also be written as
$$
{\cal D}_3=R\,{\cal D}_2=R^2{\cal D}_1\eqno(20)
$$
Using prolongation methods [14] and its generalization to the supersymmetric
systems [15] the Jacobi identity as well the compatibility of the structures
${\cal D}_1$, ${\cal D}_2$ and ${\cal D}_3$ can be checked. The second
structure (18) corresponds to the twisted $N=2$ superconformal algebra [16]
after a linear change of basis of the fields, as we will discuss at the end of
this talk.

Finally, the recursion operator defined in (19) has the form
$$
R=\pmatrix{-D\left(D^2-(D\Phi_0)\right)D^{-1}-D^{-1}\Phi_1&
2+D^{-1}\left(2\Phi_1+(D^2\Phi_0)\right)D^{-2}\cr
\noalign{\vskip 20pt}%
(D^2\Phi_1+\Phi_1D^2)D^{-1}&
D^2+(D\Phi_0)+D\Phi_1D^{-2}
}\eqno(21)
$$
and relates the conserved charges recursively as
$$
\pmatrix{{\delta H_{n+1}\over\delta\Phi_0}\cr
\noalign{\vskip 10pt}%
{\delta H_{n+1}\over\delta\Phi_1}}=R^\dagger
\pmatrix{{\delta H_{n}\over\delta\Phi_0}\cr
\noalign{\vskip 10pt}%
{\delta H_{n}\over\delta\Phi_1}}\eqno(22)
$$
\bigskip
\noindent {\bf 4. {Reductions of sTB}}
\medskip

The sTB equation reduces to many other supersymmetric integrable models such
as the sKdV and mKdV [6,10]. Let us show how the supersymmetric NLS equation
[17,5] can be obtained from the sTB system [6,9].

Defining the fermionic superfields
$$
\eqalign{Q &= \psi + \theta q\cr
{\overline Q} &= {\overline\psi} + \theta {\overline q} \cr
}
\eqno(23)
$$
from the invertible transformation
$$
\eqalign{
\Phi_0 &= - \left(D \ln (DQ)\right) +
\left(D^{-1} (\overline Q Q)\right)\cr
\Phi_1 &= - \overline Q ( DQ)\cr}
\eqno(24)
$$
we obtain from (10), after a slightly involved derivation [6], the equations
$$
\eqalign{{
\partial  Q \over \partial  t} &=
 -(D^4 Q) + 2\left(D((DQ){\overline Q})\right)Q\cr
\noalign{\vskip 4pt}%
{\partial  \overline Q \over \partial  t} &=
 (D^4 \overline Q ) - 2\left(D((D{\overline Q})Q)\right){\overline Q}\cr
}\eqno(25)
$$
These are the sNLS equations without free parameters obtained in [10] and shown
to satisfy various tests of integrability [6].

In [12] we have shown that a scalar Lax equation operator for the sNLS equation
(25) can be obtained from the Lax operator (8) of the sTB system. With the
identifications (24) we can write (8) as
$$
L=G{\widetilde L}G^{-1}\eqno(26)
$$
where
$$
\eqalign{
G=&(DQ)^{-1}\cr
{\widetilde L}=&D^2-{\overline Q}Q-(DQ)D^{-1}{\overline Q} \cr
}\eqno(27)
$$
And this makes the Lax operators, $L$ and ${\widetilde L}$, related by a gauge
transformation in the superspace. However, different from the bosonic case
[12], it is the formal adjoint of ${\widetilde L}$
$$
{\cal L}={\widetilde L}^*\eqno(28)
$$
which gives the sNLS equations (25) through the nonstandard Lax representation
$$
{\partial{\cal L}\over\partial t}=\left[{\cal L},
\left({\cal L}^2\right)_{\ge1}\right]
\eqno(29)
$$
It is possible, in principle, to obtain the bi-Hamiltonian structure for
the sNLS system using the Gelfand-Dikii method in the superspace for the scalar
Lax operator (28). However, a generalization of this method to superspace for
nonstandard systems is not yet know, But, we can derive this structure
directly from the sTB ones, through the field redefinition in (24).
This can be found in reference [9].

If we rewrite the operator ${\cal L}$, given in (28), using the supersymmetric
Leibnitz rule we get
$$
\eqalign{
{\cal L}=&-\left(D^2+{\overline Q}Q-{\overline Q}D^{-1}(DQ)\right)\cr
=&-\left(D^2+\sum_{n=-1}^\infty\Psi_n D^{-n}\right)\cr
}\eqno(30)
$$
where
$$
\eqalign{
\Psi_{-1}=&0\cr
\Psi_n=&(-1)^{[{n+1\over2}]}\,{\overline Q}(D^nQ),\quad n\ge 0}
\eqno(31)
$$
and $\Psi_{2n}$ ($\Psi_{2n+1}$) are bosonic (fermionic) superfields. In this
way $\cal L$ above has the form of the Lax operator for the sKP hierarchy and,
therefore, we can think of the sNLS system as a constrained sKP system but of
the nonstandard kind (even the sTB can be viewed as a constrained sKP system).
However, the Lax operator $\cal L$ in this case is an even parity operator [18]
and not of the usual Manin-Radul form [4]. This is a new system, namely, a
nonstandard supersymmetric KP hierarchy, and was studied in [7]. It gives a new
supersymmetric KP equation and unifies all the KP and mKP flows.

Reduction of the sTB to the sKdV and msKdV are straightforward and details
can be found in [6,9,10].
\bigskip
\noindent {\bf 5. {Nonlocal Charges}}
\medskip

As we have already pointed out the sTB equation, given by the Lax operator (8),
has conserved local charges $Q_n$ obtained from the
integer powers of the Lax operator
as in (13). Also, the sTB has a supersymmetric charge $Q$ which is local and
conserved and implements the supersymmetric transformation in (12). However,
supersymmetric integrable models also have nonlocal conserved charges. This was
first discovered in [19] for the sKdV equation and explained in the
Gelfand-Dikii formalism in [20]. For the sTB equation we also have the presence
of nonlocal charges [8] and they can be obtained from
$$
F_{2n-1\over2}=\hbox{sTr}\,L^{2n-1\over2}\qquad n=1,2,3,\dots\eqno(32)
$$
and the first ones are
$$
\eqalign{
F_{1/2}=&-\int dz\, (D^{-1}\Phi_1)\cr
F_{3/2}=&-\int dz\,\biggl[{3\over2}(D^{-1}\Phi_1)^2-\Phi_0\Phi_1-
\Bigl(D^{-1}\bigl((D\Phi_0)\Phi_1\bigl)\Bigr)\biggr]\cr
F_{5/2}=&-\int dz\,\biggl[\,{1\over6}(D^{-1}\Phi_1)^3-
\bigl(5(D^{-2}\Phi_1)\Phi_1
-2\Phi_0\Phi_1-3(D\Phi_1)-(D^{-1}\Phi_1)^2\bigr)(D\Phi_0)\cr
&\phantom{2\int dz\,\Bigl[}
+\Bigl(D^{-1}\bigl((D\Phi_1)\Phi_1+\Phi_1(D\Phi_0)^2-
(D\Phi_1)(D^2\Phi_0)\bigr)\Bigr)\biggr]\cr
}\eqno(33)
$$

These nonlocal charges are conserved and are fermionic. They reduce to the
nonlocal charges of the sKdV [19,20] if we set $\Phi_0=0$. This is natural
since we have already said that the sKdV equations is contained in the sTB
system. Also, the nonlocal charges $F_{2n-1\over2}$ are not supersymmetric
since the integrand in (33) are not local functions of superfields. Even, the
supersymmetric charge $Q$ is not supersymmetric.

We can now ask about the algebra of these charges $Q_n$, $Q$ and
$F_{2n-1\over2}$  [8]. First, we obtain
$$
\eqalign{
\{Q_n,Q_m\}_1=&0\cr
\{Q_n,F_{2m-1\over2}\}_1=&0\cr
\{Q_n,Q\}_1=&0\cr
}\eqno(34)
$$
which simply implies that these charges are conserved under any flow of the
hierarchy. The fact that $Q$ and $F_{2n-1\over2}$ are not supersymmetric is
expressed by
$$
\eqalign{
\{Q,Q\}_1=&-Q_2\cr
\{Q,F_{1/2}\}_1=&Q_1 \cr
\{Q,F_{3/2}\}_1=&{1\over2}Q_2 \cr
\{Q,F_{5/2}\}_1=&{1\over3}Q_3+{1\over24}Q_1^3 \cr
}\eqno(35)
$$
And the algebra for the lowest order nonlocal charges is
$$
\displaylines{
\hfill
\eqalign{
\{F_{1/2},F_{1/2}\}_1=&0\cr
\{F_{1/2},F_{3/2}\}_1=&Q_1 \cr
\{F_{1/2},F_{5/2}\}_1=&Q_2 \cr
}
\hfill
\eqalign{
\{F_{3/2},F_{3/2}\}_1=&2Q_2 \cr
\{F_{3/2},F_{5/2}\}_1=&{7\over3}Q_3+{7\over24}Q_1^3 \cr
\{F_{5/2},F_{5/2}\}_1=&3Q_4-{5\over 8}Q_1^2 Q_2 \cr
}
\hfill
(36)}
$$
So, the algebra of the conserved charges of the sTB system closes as a graded
algebra. The Jacobi identity is trivially satisfied since the $Q_n$'s are in
involution with all the charges. It can also be seen that the algebra
is closed with respect to the second Hamiltonian structure (18) and the
resulting algebra [8] is a sorting of shifting of (34), (35) and (36).

The cubic terms in the algebra above arise from boundary contributions when
nonlocal terms are involved. Using the following realization of the inverse
operator
$$
\bigl(\partial^{-1} f(x)\bigr) ={1\over 2}\int d y \, \epsilon(x-y)f(y)
\,,\qquad \epsilon(x)= \cases{-1,\,& $x<0$\cr \phantom{-}0,\, &$x=0$\cr
+1,\, &$x>0$\cr}
\quad \eqno(37)
$$
we can, for instance, obtain terms of the type
$$
\int dz\,(D^{-1}\Phi_1)^2\Phi_1={1\over3}\int dz\,D(D^{-1}\Phi_1)^3=
-{1\over12}Q_1^3\eqno(38)
$$
showing the origin of the nonlinear character of the algebra. The algebra
for the nonlocal charges of the sKdV equation also has cubic terms [8]. In
fact, algebras of nonlocal charges showing nonlinearity of the cubic kind are
present in various other systems and are related with Yangian structures
[21,22]. The nonlinearity in (35) and (36) appear to be redefineable to cubic
terms [8]. This is well known for the nonlinear sigma model [22]. With
appropriate redefinitions, the right hand side of the algebra (35) and (36)
appears to take the form
$$
a\,{\hat Q}_n+b\!\!\!\sum_{p+q+\ell=n}\!\!\!
{\hat Q}_p {\hat Q}_q {\hat Q}_\ell\eqno(39)
$$
which is a cubic algebra. It suggests the presence of a Yangian structure,
although the algebra for the local charges, in the present case, is involutive.

{}From (35) and (36) we see that $F_{3/2}$ has the same algebra, with the other
generators of the algebra, as the supersymmetry charge $Q$. Thus, this can be
identified with a second supersymmetry charge and this shows that the
sTB equation has in fact an $N=2$ extended supersymmetry [23].
The generators of the $N=2$ supersymmetry can be identified with
$$
{1\over3}(Q-F_{3/2})\quad\hbox{and}\quad-{1\over3}(2Q+F_{3/2})\eqno(40)
$$
Furthermore, with the linear change of variables
$$
\eqalign{
{\overline\xi}=&-{1\over2}(\psi_0'-\psi_1)\cr
\xi=&{1\over2}\psi_1\cr
}
\eqno(41)
$$
the second Hamiltonian structure (18), in terms of components,
gives the following nonvaninshing Poisson brackets [10]
$$
\eqalign{
\noalign{\vskip 3pt}%
\{J_0(x),J_0(y)\}_2=&2\delta'(x-y)\cr
\noalign{\vskip 3pt}%
\{J_0(x),J_1(y)\}_2=&(J_0\delta(x-y))'-\delta''(x-y)\cr
\noalign{\vskip 3pt}%
\{J_0(x),{\overline\xi}(y)\}_2=&-{\overline\xi}\delta(x-y)\cr
\noalign{\vskip 3pt}%
\{J_0(x),\xi(y)\}_2=&\xi\delta(x-y)\cr
\noalign{\vskip 3pt}%
\{J_1(x),J_1(y)\}_2=&J'_1\delta(x-y)+2J_1\delta'(x-y)\cr
\noalign{\vskip 3pt}%
\{J_1(x),{\overline\xi}(y)\}_2=&{\overline\xi}\delta'(x-y)\cr
\noalign{\vskip 3pt}%
\{J_1(x),\xi(y)\}_2=&\xi'\delta(x-y)+2\xi\delta'(x-y)\cr
\noalign{\vskip 3pt}%
\{{\overline\xi}(x),\xi(y)\}_2=&
-{1\over4}J_1\delta(x-y)+{1\over4}(J_0\delta(x-y))'-{1\over4}\delta''(x-y)\cr
}\eqno(42)
$$
This local algebra is nothing other than the twisted $N=2$ superconformal
algebra [16] whose bosonic limit is the Virasoro-Kac-Moody algebra for the sTB
system [10].
\bigskip
\noindent {\bf Acknowledgements}
\medskip

This work was supported in part by the U.S. Department of Energy Grant No.
DE-FG-02-91ER40685. J.C.B. would like to thank CNPq, Brazil, for
financial support.

\vfill\eject
\noindent {\bf {References}}
\bigskip

\item{1.} L.D. Faddeev and L.A. Takhtajan, ``Hamiltonian Methods in
the Theory of Solitons'' (Springer, Berlin, 1987);
A. Das, ``Integrable Models'' (World Scientific, Singapore, 1989);
L. A. Dickey, ``Soliton Equations and Hamiltonian Systems'' (World
Scientific, Singapore, 1991).

\item{2.} D. J. Gross and A. A. Midgal, Phys. Rev. Lett. {\bf 64}, 127 (1990);
D. J. Gross and A. A. Midgal, Nucl. Phys. {\bf B340}, 333 (1990); E. Br\'ezin
and V. A. Kazakov, Phys. Lett. {236B}, 144 (1990); M. Douglas and S. H.
Shenker,
Nucl. Phys. {\bf B335}, 635 (1990).

\item{3.} L. Alvarez-Gaum\'e and J. L. Man\~es, Mod. Phys. Lett. {\bf A6},
2039 (1991); L. Alvarez-Gaum\'e, H. Itoyama, J. Man\~es and A. Zadra, Int. J.
Mod. Phys. {\bf A7}, 5337 (1992).

\item{4.} Y. I. Manin and A. O. Radul, Commun. Math. Phys. {\bf 98}, 65 (1985);
P. Mathieu, J. Math. Phys. {\bf 29}, 2499 (1988).

\item{5.} J. C. Brunelli and A. Das, J. Math. Phys. {\bf 36}, 268 (1995).

\item{6.} J. C. Brunelli and A. Das, Phys. Lett. {\bf B337}, 303 (1994).

\item{7.} J. C. Brunelli and A. Das, ``A Nonstandard Supersymmetric KP
Hierarchy'', University of Rochester preprint UR-1367 (1994) (also
hep-th/9408049), to appear in the Rev. Math. Phys..

\item{8.} J. C. Brunelli and A. Das, ``Properties of Nonlocal Charges in the
Supersymmetric Two Boson Hierarchy'', University of Rochester
preprint UR-1417 (1995) (also hep-th/9504030), to appear in the Phys. Lett. B.

\item{9.} J. C. Brunelli and A. Das, ``Bi-Hamiltonian Structure of the
Supersymmetric Nonlinear Schr\"odinger Equation'', University of Rochester
preprint UR-1421 (1995) (also hep-th/9505041).

\item{10.} J. C. Brunelli and A. Das, ``Supersymmetric Two Boson Equation, Its
Reductions and the Nonstandard Supersymmetric KP Hierarchy'',
University of Rochester preprint UR-1422 (1995) (also hep-th/9505093),
to appear in the Int. J. Mod. Phys. A.

\item{11.} B.A. Kupershmidt, Commun. Math. Phys. {\bf 99}, 51 (1985).

\item{12.} H. Aratyn, L.A. Ferreira, J.F. Gomes and A.H. Zimerman, Nucl.
Phys. {\bf B402}, 85 (1993); L. Bonora and C.S. Xiong,
Int. J. Mod. Phys. {\bf A8}, 2973 (1993);
M. Freeman and P. West, Phys. Lett. {\bf 295B}, 59 (1992);
J. Schiff, ``The Nonlinear Schr\"odinger Equation and
Conserved Quantities in the Deformed Parafermion and SL(2,{\bf R})/U(1)
Coset Models'', Princeton preprint IASSNS-HEP-92/57 (1992)
(also hep-th/9210029).

\item{13.} J. C. Brunelli, A. Das and W.-J. Huang, Mod. Phys. Lett. {\bf 9A},
2147 (1994).

\item{14.} P. J. Olver , ``Applications of Lie Groups to Differential
Equations'', Graduate Texts in Mathematics, Vol. 107 (Springer, New York,
1986).

\item{15.} P. Mathieu, Lett. Math. Phys. {\bf 16}, 199 (1988).

\item{16.} E. Witten, Commun. Math. Phys. {\bf 117}, 353 (1988); {\bf 118},
411 (1988); Nucl. Phys. {\bf B340} 281 (1990); T. Eguchi and S. Yang, Mod.
Phys.
Lett. {\bf A4}, 1693 (1990).

\item{17.} G.H.M. Roelofs and P.H.M. Kersten, J. Math. Phys. {\bf 33}, 2185
(1992).

\item{18.} J. M. Figueroa-O'Farrill, J. Mas and E. Ramos, Rev. Math. Phys.
{\bf 3}, 479 (1991);
F. Yu, Nucl. Phys. {\bf B375}, 173 (1992);
J. Barcelos-Neto, S. Ghosh and S. Roy, J. Math. Phys. {\bf 36}, 258 (1995).

\item{19.} P. H. M. Kersten, Phys. Lett. {\bf A134}, 25 (1988).

\item{20.} P. Dargis and P. Mathieu, Phys. Lett. {\bf A176}, 67 (1993).

\item{21.} D. Bernard and A. LeClair, Commun. Math. Phys. {\bf 142}, 99 (1989);
D. Bernard, ``An Introduction to Yangian Symmetries'', in Integrable
Quantum Field Theories, ed. L. Bonora et al., NATO ASI Series B: Physics vol.
310 (Plenum Press, New York, 1993);
J. Barcelos-Neto, A. Das, J. Maharana, Z. Phys. {\bf 30C}, 401 (1986);
N. J. Mackay, Phys. Lett. {\bf B281}, 90 (1992);
T. Curtright and C. Zachos, Nucl. Phys. {\bf B402}, 604 (1993).

\item{22.} E. Abdalla, M. C. B. Abdalla, J. C. Brunelli and A. Zadra, Commun.
Math. Phys. {\bf 166}, 379 (1994).

\item{23.} S. Krivonos and A. Sorin, ``The Minimal $N=2$ Superextension of the
NLS Equation'', preprint JINR E2-95-172 (also hep-th/9504084);
S. Krivonos, A. Sorin and F. Toppan, ``On the Super-NLS Equation and
its Relation with $N=2$ Super-KdV within Coset Approach'', preprint
JINR E2-95-185 (also hep-th/9504138).

\end